\documentstyle[12pt,axodraw]{article}
\begin{document}
\newcommand{\bfig}{\begin{center}\begin{picture}}
\newcommand{\efig}[1]{\end{picture}\\{\small #1}\end{center}}
\newcommand{\flin}[2]{\ArrowLine(#1)(#2)}
\newcommand{\wlin}[2]{\DashLine(#1)(#2){2.5}}
\newcommand{\zlin}[2]{\DashLine(#1)(#2){5}}
\newcommand{\glin}[3]{\Photon(#1)(#2){2}{#3}}
\newcommand{\lin}[2]{\Line(#1)(#2)}

\newcommand{\sof}{\SetOffset}
\newcommand{\bmip}[2]{\begin{minipage}[t]{#1pt}\bfig(#1,#2)}
\newcommand{\emip}[1]{\efig{#1}\end{minipage}}
\newcommand{\putk}[2]{\Text(#1)[r]{$p_{#2}$}}
\newcommand{\putp}[2]{\Text(#1)[l]{$p_{#2}$}}
\newcommand{\bq}{\begin{equation}}
\newcommand{\eq}{\end{equation}}
\newcommand{\bqa}{\begin{eqnarray}}
\newcommand{\eqa}{\end{eqnarray}}
\newcommand{\nl}{\nonumber \\}
\newcommand{\eqn}[1]{eq. (\ref{#1})}
\newcommand{\eqs}[1]{eqs. (\ref{#1})}
\newcommand{\ibidem}{{\it ibidem\/},}
\newcommand{\vpb}{}
\newcommand{\p}[1]{{\scriptstyle{\,(#1)}}}
\newcommand{\gev}{\mbox{GeV}}
\newcommand{\mev}{\mbox{MeV}}
\title{
\vspace{-4cm}
\begin{flushright}
{\large ~~~ } \\
\end{flushright}
\vspace{2.7cm}
Four-fermion production in electron-positron collisions
with {\tt NEXTCALIBUR}
     }
\author{
        {\bf F. A. Berends}\\
        Instituut Lorentz, University of Leiden, P. O. Box 9506,\\
        2300 RA Leiden, The Netherlands\\\\
        {\bf C.~G.~Papadopoulos}\\
        Institute of Nuclear Physics, NCRS 'Democritos',\\
        15310 Athens, Greece\\\\
          and\\\\
        {\bf R.~Pittau}\\
        Dipartimento di Fisica Teorica, 
Universit\`a di Torino, Italy\\
INFN, Sezione di Torino, Italy}
\maketitle
\thispagestyle{empty}
\begin{abstract}
\end{abstract}
We introduce a new, fully massive, Monte Carlo program
to compute all four-fermion processes in $e^+e^-$ collisions,
including Higgs production. We outline our strategy 
for the matrix element evaluation, the phase space generation 
and the implementation of the leading higher order effects, 
and show, where available, comparisons with existing results.

\vspace{1.3cm}

\noindent Pacs: 11.15.-q, 13.10.-q, 13.38-b, 13.40,-f, 14.70.Fm

\clearpage
\section{Introduction}
Since the beginning of LEP2 four-fermion Physics underwent an 
extensive  development.
From an experimental point of view, new processes and effects, 
not included in the 1996 LEP2 Yellow Report \cite{lep2}, became relevant. 
All codes used so far have to be upgraded and extended with
new features in order to match the improved experimental needs, 
especially in view of the final LEP2 analysis.

Although the needed improvements of the various event
generators in use at LEP2 may differ from one code
to another 
there often are three obvious steps to be taken, namely
\begin{itemize}
\item Including fermion masses, instead of neglecting them.
\item Taking into account scale-dependent corrections
      and higher order contributions related
      to unstable particles.
\item Improving the treatment of the QED radiation.
\end{itemize}

In the first place,
fermion masses are relevant both for Higgs production and
for single-$W$ \cite{singlew} or $\gamma\,\gamma$ dominated processes 
\cite{gg}, when electrons are scattered in the very forward region.

Secondly, the correct scales have to be taken into account for 
processes dominated by quasi-real $t$-channel photons
(single-$W$ production) or $s$-channel photons with low virtuality
($Z \gamma^\ast$ processes), where
tools adequate for studying high energy dominated processes, 
such as $W^+W^-$ production, fail in describing the data.

Finally, it would be desirable to generate a realistic non-vanishing 
$p_t$ distribution for the radiated photons.

These improvements should be made while, 
 at the same time, the Monte Carlo program should be kept 
general enough to deal with all processes 
in all possible kinematical configurations. It is this
requirement of having all processes and all kinematical
configurations in one program, which is the challenge. 

This paper paves the way to replace an existing  
code, {\tt EXCALIBUR} \cite{exca}, 
which describes all final states, by a new {\tt FORTRAN} program, 
{\tt NEXTCALIBUR}, which contains the above list
of improvements. The demand of massive fermions
leads us to a new type of matrix element calculation
and severe requirements on the phase space generation.
This means in practice a new program, since the
old program made essential use of massless fermions.

The structure of the work is as follows.
In section 2, we describe the strategy for the massive matrix 
element evaluation and for the Monte Carlo integration
in {\tt NEXTCALIBUR}.
In section 3 we present the treatment of the leading 
higher order contributions, mainly QED radiation and 
running of $\alpha_{QED}$. Subsequently, 
in section 4, we show numbers and
comparisons with other results in the literature.
Finally, the last section is devoted to our conclusions. 
\section{The strategy}
We solved the problem of the complexity in the massive matrix
element evaluation by using {\tt HELAC}, a {\tt FORTRAN} package
for helicity amplitudes computation based on Dyson-Schwinger
equations, as described in ref.~\cite{costas},
to which we refer for more details. 
Here we only point out that, 
in the case of
massless particles, the algorithm is simplified from the beginning, 
by avoiding the computation of helicity amplitudes that are known to be zero.
A very fast computation of the cross section can therefore be obtained. 
In table 1 we report comparisons in speed  between 
{\tt NEXTCALIBUR} (fully massive) and {\tt EXCALIBUR} (massless)
for three processes.

\vspace{0.5cm}

\begin{center}
\begin{tabular}{|l||c|c|} \hline 
Process        & {\tt NEXTCALIBUR}  
               & {\tt ~EXCALIBUR~}              
      \\ \hline \hline
$e^-~\bar \nu_e~\nu_e~e^+$ & 34.6 (16) &12.3 (8) \\ \hline              
$e^-~\bar \nu_e~u~\bar d $ & 35.0 (32) & 5.4 (3) \\ \hline               
$\mu^+~\mu^-~b~\bar b    $ & 62.0 (64) & 6.1 (8) \\ \hline              
\end{tabular}
\end{center}
{\em Table 1: CPU time comparison between {\tt NEXTCALIBUR} and
{\tt EXCALIBUR} in seconds/1000 points. In parenthesis we
show the number of non vanishing helicity configurations.
The speed difference is mainly due to the increased number of 
contributing helicities.}

\vspace{0.4cm}

As for the phase space generation, we used the same multi-channel 
self-adjusting approach used in {\tt EXCALIBUR}. 
Namely, we wrote a set
of kinematical channels, each of them taking into account 
a different peaking structure of the integrand. 
The difference with the {\tt EXCALIBUR} channels  
is that now all fermion masses are taken into account. 
Furthermore the leading kinematical structures
for Higgs production have been added (see fig. 1a). 

More details can be found elsewhere \cite{future}. We just mention here 
that the most complicated channel is the double multi-peripheral 
configuration given in fig. 1b, because three $t$-channel 
like singularities have to be mapped simultaneously: two of them due to 
quasi-real photons, and the third one induced by the fermion propagator.
\begin{figure}
\begin{center}
\begin{picture}(100,105)(0,-50)
\sof(-80,0)
\flin{-15,15}{0,-5} \flin{0,-5}{-15,-25}
\glin{0,-5}{30,-5}{5}
\glin{30,20}{30,-5}{4}
\wlin{30,-5}{30,-30}
\flin{45,0}{30,20} \flin{30,20}{45,40}
\flin{30,-30}{45,-10} \flin{45,-50}{30,-30}
\Text(-20,-30)[r]{\small 2} \Text(-20,20)[r]{\small 1}
\Text(50,40)[l]{\small 3} \Text(50,0)[l]{\small 4}
\Text(50,-10)[l]{\small 5} \Text(50,-50)[l]{\small 6}
\Text(-20,40)[r]{(a)}
\Text(25,12)[r]{$Z$}
\Text(25,-23)[r]{$H$}
\Text(12,5)[r]{$Z$}
\sof(20,0)
\flin{0,40}{11,40} \lin{11,40}{33,40} \flin{33,40}{45,40}
\glin{13,40}{13,-5}{5}
\glin{13,-5}{13,-50}{5}
\wlin{13,-5}{43,-5}
\flin{43,-5}{55,15} \flin{55,-25}{43,-5}
\flin{11,-50}{0,-50} \lin{33,-50}{11,-50} \flin{45,-50}{33,-50}
\Text(-5,-50)[r]{\small 2} \Text(-5,40)[r]{\small 1}
\Text(50,40)[l]{\small 3} \Text(60,20)[l]{\small 4}
\Text(60,-30)[l]{\small 5} \Text(50,-50)[l]{\small 6}
\Text(7,15)[r]{W,Z}
\Text(7,-25)[r]{W,Z}
\Text(30,0)[b]{H}
\sof(150,0)
\flin{0,40}{11,40} \lin{11,40}{33,40} \flin{33,40}{45,40}
\glin{22,40}{22,10}{4}
\flin{22,10}{45,10} \flin{22,-20}{22,10} \flin{45,-20}{22,-20}
\glin{22,-50}{22,-20}{4}
\flin{11,-50}{0,-50} \lin{33,-50}{11,-50} \flin{45,-50}{33,-50}
\Text(-5,-50)[r]{\small 2} \Text(-5,40)[r]{\small 1}
\Text(50,40)[l]{\small 3} \Text(50,10)[l]{\small 4}
\Text(50,-20)[l]{\small 5} \Text(50,-50)[l]{\small 6}
\Text(15,20)[r]{$\gamma$}
\Text(15,-30)[r]{$\gamma$}
\Text(-20,40)[r]{(b)}
\end{picture}
\end{center}
{\em Fig 1: Higgs production {\rm (a)} and 
double multi-peripheral {\rm (b)}  
kinematical channels in {\tt NEXTCALIBUR}.}
\end{figure}

\section{Implementing higher order contributions}
 A first source of numerically important higher order 
contributions comes from the widths of the unstable bosons,
that have to be included without breaking gauge invariance.
 Our approach is to use complex bosonic masses everywhere, 
also in the definition of the weak mixing angle. 
This obeys all relevant Ward Identities \cite{denner} 
and has been shown to be a very good approximation even in the forward region 
for $t$-channel dominated processes \cite{sandro}.

A second source of large higher order effects is the QED radiation. 
A very common solution is using the Structure Function formalism, 
namely a convolution of the Born cross section together with 
QED Initial State Radiators \cite{isr}.
Such a strategy is implemented in most of the programs used for 
the analysis of the LEP2 data \cite{isr1} and accurately reproduces
the inclusive four-fermion cross sections, at least for $s$-channel dominated
processes. 

  Recent studies have shown, by inspection with the soft limit 
of exact calculations, that the Structure Function formalism can still 
be used for $t$-channel dominated processes, provided the scale $q^2$ 
of the radiators is chosen to be of the order of the virtuality 
of the exchanged $t$-channel photons. 
Our approach to the QED corrections is therefore using Initial State 
Structure Functions with a proper choice of $q^2$ 
for each of the two incoming legs
\footnote{Presently, Final State radiation is neglected, but it
can be included with analogous techniques.}.

As to the scale $q^2$, $s$ should be taken
for $s$-channel dominated processes, while, when a process is
dominated by small $t$ exchanges and $-t$
is much smaller than $s$, the scale is related to $t$. 
This is e.g. the case in small angle Bhabha scattering \cite{sbhab}
and the proper scale is chosen as the one which reproduces
roughly the exact first order QED correction, which is known
for Bhabha scattering. A similar procedure now also exists
for the multi-peripheral two photon process \cite{isr2}, since
an exact first order calculation is also available \cite{gge}.
In these $t$-channel dominated processes it is important to know whether
a cross-section with angular cuts is wanted, since then the
$t$-related scale will increase and the QED corrections as well.
When no exact first order calculations are available the
scale occurring in the first order soft corrections is
also used as guideline to determine $q^2$ \cite{isr2,isr3}.

 In {\tt NEXTCALIBUR} the choice of the
scale is performed automatically by the program, event by event, according 
to the selected final state (see Table 2).

\vspace{0.5cm}

\begin{center}
\begin{tabular}{|l||c|c|} \hline 
Final State          & $q^2_-$  & $q^2_+$ \\ \hline \hline
No $e^\pm$ & $s$     & $s$     \\ \hline              
1  $e^-$   & $|t_-|$ & $s$     \\ \hline               
1  $e^+$   & $s$     & $|t_+|$ \\ \hline               
1  $e^-$ and 1  $e^+$  & $|t_-|$    & $|t_+|$ \\ \hline               
2  $e^-$ and 2  $e^+$  & min($|t_-|$)    & min($|t_+|$) \\ \hline
\end{tabular}
\end{center}
{\em Table 2: The choice of the QED scale in {\tt NEXTCALIBUR}.
$q^2_\pm$ are the scales of the incoming $e^\pm$ while
$t_\pm$ represent the $t$-channel invariants obtained
by combining initial and final state $e^\pm$ momenta.
When two combinations are possible, as in the last entry of the table, 
that one with the minimum value of $|t|$ is chosen, event by event.}
\vspace{0.4cm}

  However, as mentioned before, the choice of the appropriate $q^2$ is not 
the only required improvement to the treatment of the QED radiation.
The increasing precision of the collected LEP2 data also requires
the knowledge of the $p_t$ spectrum of the emitted photons.
 We therefore use a particular form of $p_t$ dependent 
Structure Functions \cite{ptsf}, derived, at the first leading logarithmic
order, for small values of $p_t$. 

In practice, we replace the quantity 
$$\ln(\frac{q^2}{m^2_e})~~~~{\rm by}~~~~  
\frac{1}{1-c_i+2\frac{m^2_e}{q^2}}$$ in the 
strictly collinear Structure Function for 
the $i^{th}$ incoming particle, by explicitly generating
$c_1$ and $c_2$, the cosines (in the laboratory frame)
of the emitted photons with respect to the incoming particles.
Once $c_{1,2}$ are generated, together with the energy
fractions $x_{1,2}$ and the azimuthal angles $\phi_{1,2}$, 
the momenta of two ISR photons are known.
The four-fermion event is then generated in the c.m.s.
of the incoming particles {\em after} QED radiation, and then
boosted back to the laboratory frame. 

We also take into account non leading terms 
with the substitution \cite{ptsf1}
$$\ln(\frac{q^2}{m^2_e})-1~~\to~~~~~ 
\frac{1}{1-c_i+2\frac{m^2_e}{q^2}} -2 \frac{m^2_e}{q^2}
\frac{1}{(1-c_i+2\frac{m^2_e}{q^2})^2}.$$

The above choice ensures that the residue of the 
soft-photon pole gets proportional to $\ln(\frac{q^2}{m^2_e})-1$,
after integration over $c_i$. The inclusive QED result 
is therefore exactly recovered, after integrating over
the $p_t$ spectrum and, at the same time, 
the pattern of the photon radiation is exact for small $p_t$ values.
Notice that the $p_t$ spectrum is controlled by the
same scale $q^2$ used in the strictly collinear Structure Functions, namely
an $s$-channel scale for $s$-channel processes and a $t$-channel scale
for $t$-channel dominated final states. 
The radiation pattern is therefore different in the two situations,
as one naively expects.
The presence of such a scale can also be thought as an extra handle
to tune our Monte Carlo predictions to the data, to get a satisfactory 
description of the radiation. 

Another problem, in presence of low $t$-channel scales, is that 
high energy renormalization schemes, such as the $G_F$ scheme, fail 
in describing the data, because of the running 
of $\alpha_{QED}$.

A possible solution is the Fermion-Loop approach of refs. 
\cite{passa1}-\cite{passa3}, where all fermion corrections  
are consistently included by introducing 
running couplings $g(s)$ and $e(s)$ and re-summed bosonic propagators.

In presence of the $W W \gamma$ vertex, the above ingredients 
are not sufficient to ensure gauge invariance, because loop 
mediated vertices have to be consistently included. 
On the contrary, when no $W W \gamma$ vertex is present, 
the neutral gauge boson vertices, induced by the 
fermion loop contributions, are separately gauge invariant \cite{passa2}.

 Instead of explicitly including the loop vertices, we follow 
a {\em Modified Fermion-Loop} approach. 
Namely, we neglect the separately 
gauge invariant neutral boson vertices, and include 
only the part of the $W W \gamma $ loop function necessary 
to renormalize the bare $W W \gamma$ vertex and 
to insure the $U(1)$ gauge invariance
\footnote{When using {\tt NEXTCALIBUR} in this {\em running 
coupling mode}, $U(1)$ is preserved but $SU(2)$ is, in general, violated. 
The numerical effects of this are expected to be small 
at LEP2 energies. The implementation in the code of the fully
$SU(2) \times U(1)$ gauge invariant solution of ref.~\cite{bbc} 
is under way.}. 
Our procedure is as follows. Besides running couplings,
we use bosonic propagators
\bqa
&&P_w^{\mu\nu}(s)= \left(s-M^2_w(s)\right)^{-1}
 \left(g_{\mu\nu}-\frac{p_\mu p_\nu}{M^2_w(s)}\right)
 \nonumber \\
&&P_z^{\mu\nu}(s)= \left(s-M^2_z(s)\right)^{-1} 
 \left(g_{\mu\nu}-\frac{p_\mu p_\nu}{M^2_z(s)}\right)
\nonumber 
\eqa
with running boson masses defined as
\bqa
&& M^2_w(s)= \mu_w\frac{g^2(s)}{g^2(\mu_w)}
-g^2(s) [T_W(s)-T_W(\mu_w)]\,\nonumber \\
&& M^2_z(s)= \mu_z \frac{g^2(s)}{c^2_\theta(s)} 
\frac{c^2_\theta(\mu_z)}{g^2(\mu_z)}
-\frac{g^2(s)}{c^2_\theta(s)}[T_Z(s)-T_Z(\mu_z)]\,. \nonumber
\eqa
$T_{W,Z}(s)$ are contributions due to the top quark,
$\mu_{w,z}$ the complex poles of the propagators (one can
take, for instance, $\mu_{w,z}= M^2_{w,z}-i\Gamma_{w,z} M_{w,z}$) and
$$ s^2_\theta(s)= \frac{e^2(s)}{g^2(s)}\,,~c^2_\theta(s)= 1-s^2_\theta(s)\,.$$
The explicit form of the running functions $e^2(s),~g^2(s)$ and
$T_{W,Z}(s)$ can be found in ref. \cite{passa2}. 

The leading contributions are in the real part of the running couplings
therefore we take only the real part of them. This also means that one can
replace, in the above formulae, $g^2(\mu_{w,z}) \to g^2 (M^2_{w,z})$,
$c^2_\theta(\mu_z) \to c^2_\theta(M^2_z)$ and 
also $T_{W,Z}(\mu_{w,z}) \to T_{W,Z}(M^2_{w,z})$.

When the $W W \gamma$ coupling is present, we 
introduce, in addition, the following effective three gauge boson vertex 
\begin{center}
\begin{picture}(100,100)(50,-50)
\glin{0,0}{30,0}{4}
\glin{30,0}{60,20}{4}
\glin{30,0}{60,-20}{4}
\Vertex(30,0){2.5}
\LongArrow(5,7)(20,7)
\LongArrow(50,25)(40,17)
\LongArrow(50,-25)(40,-17)
\Text(-3,0)[r]{\small $\gamma_\mu$}
\Text(63,20)[bl]{\small $W^+_\nu$}
\Text(63,-20)[tl]{\small$W^-_\rho$}
\Text(7,12)[bl]{\small $p$}
\Text(44,25)[br]{\small  $p_+$}
\Text(44,-23)[tr]{\small $p_-$}
\Text(80,0)[l]{$= i\,e(s)V_{\mu \nu \rho}$}
\end{picture}
\end{center}
with $s = p^2\,,~~s^+= p^2_+\,,~~s^-= p^2_-$ and
\bqa
V_{\mu \nu \rho} &=& 
 g_{\mu\nu}  (p   -p_+)_\rho
+g_{\nu\rho} (p_+ -p_-)_\mu\,(1+\delta_V)
+g_{\rho\mu} (p_- -p  )_\nu  \nonumber \\
&+&\frac{(p_+ -p_-)_\mu}{s^- - s^+}\left[
 \left(\frac{g(s^-)}{g(s^+)}-1 \right)\,p_{+ \nu}p_{+ \rho}
-\left(\frac{g(s^+)}{g(s^-)}-1 \right)\,p_{- \nu}p_{- \rho}
                             \right] \nonumber \\
\delta_V&=&  \frac{1}{g(s^+) g(s^-) (s^- - s^+)} 
  \left[ g^2(s^+) g^2(s^-)\,[ T_W(s^-)- T_W(s^+) ]  \right. 
\nonumber \\
     &+& \left. [g(s^+)- g(s^-)]\,[ s^- g(s^+) + s^+ g(s^-) ] 
\right]\,. 
\eqa
It is the easy to see that, with the above choice for $V_{\mu \nu \rho}$,
the $U(1)$ gauge invariance - namely current conservation -
is preserved, even in presence of complex 
masses and running couplings, also with massive final state fermions.

From eq. (1), one deduces at least two effective ways 
to preserve $U(1)$. One can either
compute $g(s)$ at a fixed scale (for example always with $s= M_W^2$),
while keeping only the running of $e(s)$, or let all the couplings 
run at the proper scale.
 With the first choice the modification of the three gauge boson 
vertex is kept minimal (but the leading running effects included).
With the second choice everything runs, but a heavier 
modification of the Feynman rules is required.
At this point one should not forget that our approach is an effective one, 
the goodness of which can be judged only by comparing with the exact 
calculation of ref.~\cite{passa1}.
We found that the second choice gives a better agreement
for leptonic single-$W$ final states, while the first one 
is closer to the exact result in the hadronic case, which
is phenomenologically more relevant. Therefore, we adopted this
first option as our default implementation in {\tt NEXTCALIBUR}.

 We want to stress once more that the outlined solution 
is flexible enough to deal with any four-fermion final state, 
whenever small scales dominate.  For example, once the given formulae 
are implemented in the Monte Carlo, the correct running 
of $\alpha_{QED}$ is taken into account also for $s$-channel
processes as $Z \gamma^\ast$ production.

\section{Numerical results}
In tables 3 and 4, we show the total cross sections for the processes 
$e^+ e^- \to e^+ e^- \mu^+ \mu^-$ and
$e^+ e^- \to e^+ e^- e^+ e^-$. Where available, we compare our predictions
with the QED numbers published in ref. \cite{gg}.

\vspace{0.3cm}

\begin{center}
\begin{tabular}{|l||c|c|} \hline 
$\sqrt{s}$       & {\tt BDK}  
                 & {\tt NEXTCALIBUR}              
      \\ \hline \hline
 20      &          98.9 $\pm$ 0.6 &   99.20 $\pm$ 0.98  \\ \hline          
 35      &         131.4 $\pm$ 2.2 &  131.03 $\pm$ 0.88  \\ \hline
 50      &         154.4 $\pm$ 0.9 &  152.33 $\pm$ 0.83  \\ \hline
100      &         205.9 $\pm$ 1.2 &  204.17 $\pm$ 1.73  \\ \hline
200      &            ---          &  263.50 $\pm$ 1.31  \\ \hline 
200 (all)&            ---          &  265.58 $\pm$ 1.44  \\ \hline  
\end{tabular}
\end{center}
{\em Table 3: $\sigma_{tot}$ (in nb) for the process
$e^+ e^- \to e^+ e^- \mu^+ \mu^-$. Only QED diagrams, 
except in the last entry.}

\vspace{0.5cm}

\noindent {\tt NEXTCALIBUR} contains all electroweak 
diagrams, and can therefore be used to compute 
the electroweak background to the above 
$\gamma\,\gamma$ processes.
By looking at the last entry of the tables, the latter is found to be less
than 1 \% at LEP2 energies, at least for totally inclusive quantities.

   All numbers have been produced at the Born level, but ISR and running
$\alpha_{QED}$ can be 
included as described in the previous section. Here we mainly want 
to demonstrate the ability of the program to cover all phase-space regions,
without loosing efficiency in the event generation.

At the moment, in order to get the necessary numerical accuracy, 
we run the program in quadruple precision. 
{\tt NEXTCALIBUR} has been written in such a way that 
switching from double to quadruple precision 
simply implies adding a flag at the compilation time.
However, this option is really necessary only when two or 
more electrons are allowed in the very forward direction. 
For all the other kinematical configurations, with at most one
electron lost in the beam pipe, double precision is sufficient.
A version of the program using double precision in
all possible situations is currently under study.
\begin{figure}
\begin{center}
\begin{tabular}{|l||c|c|} \hline 
$\sqrt{s}$       & {\tt BDK}  
                 & {\tt NEXTCALIBUR}              
      \\ \hline \hline
 20      & 0.920         $\pm$ .011 &0.905   $\pm$ .011  \\ \hline          
 35      & 1.070         $\pm$ .015 &1.079   $\pm$ .014  \\ \hline
 50      & 1.233         $\pm$ .018 &1.214   $\pm$ .016  \\ \hline
100      & 1.459         $\pm$ .025 &1.485   $\pm$ .020  \\ \hline
200      &              ---         &1.776   $\pm$ .019  \\ \hline 
200 (all)&              ---         &1.787   $\pm$ .030  \\ \hline  
\end{tabular}
\end{center}
{\em Table 4: $\sigma_{tot}$ (in nb $ \times 10^7$) for the process
$e^+ e^- \to e^+ e^- e^+ e^-$. Only QED diagrams, 
except in the last entry.}
\end{figure}

In tables 5, 6 and 7 we show, as a second example, single-$W$ numbers produced
with our Modified Fermion-Loop approach, as discussed in the previous
section.
Comparisons are made with the exact Fermion-Loop calculation of ref. 
\cite{passa1}.
The results of the complete Fermion-Loop are reproduced 
at 2\% accuracy for both leptonic and hadronic 
single-$W$ final states.

  It should also be noted that, 
when neglecting Fermion-Loop corrections, 
one can directly compare {\tt NEXTCALIBUR} with other massive
Monte Carlo's and one finds excellent agreement for
single-$W$ production in the whole phase space \cite{wshop}.

In Figs. 2 and 3, we show, as an illustrative example, 
two distributions for the most energetic radiated photon in the processes
$e^+ e^- \to \mu^- \bar \nu_\mu u \bar d (\gamma)$, as predicted
by ${\tt NEXTCALIBUR}$. Only ISR photons are taken into
account, $\sqrt{s}= 200$ GeV, $|\cos \theta_\mu| < 0.985 $,
$ E_\mu > 5$ GeV and $M(u \bar d) > 10 $ GeV.
With the same set of events, by using as separation cuts
for the emitted photons $E_\gamma > 1$ GeV 
and $|\cos \theta_\gamma| <$ 0.985, we found the following values for
the total, the non-radiative, the single-radiative (1 generated $\gamma$) 
and the double-radiative (2 generated $\gamma$'s) cross sections:
\begin{center}
\begin{tabular}{rr} 
 $\sigma_{tot}  = 0.61727 \pm 0.0059~{\rm pb}$ &
 $\sigma_{n-rad}= 0.57819 \pm 0.0058~{\rm pb}$ \\
 $\sigma_{s-rad}= 0.03854 \pm 0.0016~{\rm pb}$ &
 $\sigma_{d-rad}= 0.00054 \pm 0.0002~{\rm pb}$ 
\end{tabular} 
\end{center}

\vspace{0.5cm}

\begin{center}
\begin{tabular}{|l||c|c|c|} \hline 
$d\sigma/d\theta_e$     & MFL & EFL & MFL/EFL $-$ 1 (percent)  
         \\ \hline \hline
$0.0^\circ \div 0.1^\circ$  & 0.45062(70)  &  0.44784 & +0.62\\ \hline 
$0.1^\circ \div 0.2^\circ$  & 0.06636(28)  &  0.06605 & +0.47\\ \hline 
$0.2^\circ \div 0.3^\circ$  & 0.03848(21)  &  0.03860 & -0.31\\ \hline 
$0.3^\circ \div 0.4^\circ$  & 0.02726(18)  &  0.02736 & -0.37\\ \hline \hline
$\sigma_{tot} $             & 83.26(9)     &  83.28(6)& -0.02\\ \hline
\end{tabular}                 
\end{center}
{\em Table 5: $d\sigma/d\theta_e$ [pb/degrees] and $\sigma_{tot}$ [fb] 
for the process $e^+ e^- \to e^- \bar \nu_e u \bar d$. The first column
is our Modified Fermion-Loop, the second one is the exact Fermion-Loop 
of ref. \cite{passa1}. $\sqrt{s}= 183$ GeV, $|\cos \theta_e| > 0.997$, 
$M(u\bar d) > 45$ GeV. QED radiation not included.
The number in parenthesis shows, when available, the integration 
error on the last digits.}

\vspace{2.5cm}

\begin{center}
\begin{tabular}{|l||c|c|c|} \hline 
 $d\sigma/d\theta_e$        & MFL & EFL & MFL/EFL $-$ 1 (percent)  
         \\ \hline \hline
$0.0^\circ \div 0.1^\circ$  & 0.13218(26) & 0.13448 &-1.7 \\ \hline 
$0.1^\circ \div 0.2^\circ$  & 0.01997(10) & 0.02031 &-1.7 \\ \hline 
$0.2^\circ \div 0.3^\circ$  & 0.01171(8)  & 0.01194 &-1.9 \\ \hline 
$0.3^\circ \div 0.4^\circ$  & 0.00838(6)  & 0.00851 &-1.5 \\ \hline \hline
$\sigma_{tot} $             & 25.01(3)    & 25.53(4)&-2.0 \\ \hline
\end{tabular}                 
\end{center}
{\em Table 6: $d\sigma/d\theta_e$ [pb/degrees] and $\sigma_{tot}$ [fb] 
for the process $e^+ e^- \to e^- \bar \nu_e \nu_\mu \mu^+$. The first column
is our Modified Fermion-Loop, the second one is the exact Fermion-Loop 
of ref. \cite{passa1}. $\sqrt{s}= 183$ GeV, $|\cos \theta_e| > 0.997$,
$|\cos \theta_\mu| < 0.95$ and  $E_\mu > 15$ GeV. QED radiation not included.
The number in parenthesis shows, when available, the integration 
error on the last digits.}

\clearpage

\begin{center}
\begin{tabular}{|l||c|c|c|} \hline 
$d\sigma/d\theta_e$     & MFL & EFL & MFL/EFL $-$ 1 (percent)  
         \\ \hline \hline
$0.0^\circ \div 0.1^\circ$  & 0.62694(98)& 0.62357  &+0.54 \\ \hline 
$0.1^\circ \div 0.2^\circ$  & 0.08850(38)& 0.08798  &+0.59 \\ \hline 
$0.2^\circ \div 0.3^\circ$  & 0.05100(30)& 0.05141  &-0.80 \\ \hline 
$0.3^\circ \div 0.4^\circ$  & 0.03672(25)& 0.03646  &+0.71 \\ \hline \hline
$\sigma_{tot} $             & 113.73 (13)& 113.67(8)&+0.05 \\ \hline
\end{tabular}                 
\end{center}
{\em Table 7: $d\sigma/d\theta_e$ [pb/degrees] and $\sigma_{tot}$ [fb] 
for the process $e^+ e^- \to e^- \bar \nu_e u \bar d$. The first column
is our Modified Fermion-Loop, the second one is the exact Fermion-Loop 
of ref. \cite{passa1}. $\sqrt{s}= 200$ GeV, $|\cos \theta_e| > 0.997$, 
$M(u\bar d) > 45$ GeV. QED radiation not included.
The number in parenthesis shows, when available, the integration error 
on the last digits.}

\begin{center}
\begin{picture}(400,275)(-50,0)
\LinAxis(0,50)(300,50)(10,2,5,0,1.5)
\LinAxis(0,250)(300,250)(10,2,-5,0,1.5)
\LogAxis(0,50)(0,250)(4,-5,0,1.5)
\LogAxis(300,50)(300,250)(4,5,0,1.5)
\Text(0  ,40)[t]{$-1$}
\Text(150,40)[t]{$0$}
\Text(300,40)[t]{$1$}
\Text(-27,100)[l]{$10^{-1}$}
\Text(-27,150)[l]{$10^{0 }$}
\Text(-27,200)[l]{$10^{1 }$}
\Text(-50,227)[l]{$\frac{1}{\sigma} \frac{d\sigma}{d \cos \theta_\gamma}$}
\Text(150,20)[t]{$\cos \theta_\gamma$}
\Line(   0.0, 215.4)(   3.0, 215.4)
\Line(   3.0, 142.7)(   6.0, 142.7)
\Line(   6.0, 130.9)(   9.0, 130.9)
\Line(   9.0, 123.8)(  12.0, 123.8)
\Line(  12.0, 119.0)(  15.0, 119.0)
\Line(  15.0, 114.9)(  18.0, 114.9)
\Line(  18.0, 111.2)(  21.0, 111.2)
\Line(  21.0, 107.4)(  24.0, 107.4)
\Line(  24.0, 106.8)(  27.0, 106.8)
\Line(  27.0, 103.4)(  30.0, 103.4)
\Line(  30.0, 101.2)(  33.0, 101.2)
\Line(  33.0,  99.9)(  36.0,  99.9)
\Line(  36.0,  98.8)(  39.0,  98.8)
\Line(  39.0,  95.8)(  42.0,  95.8)
\Line(  42.0,  95.0)(  45.0,  95.0)
\Line(  45.0,  94.4)(  48.0,  94.4)
\Line(  48.0,  93.5)(  51.0,  93.5)
\Line(  51.0,  92.5)(  54.0,  92.5)
\Line(  54.0,  91.6)(  57.0,  91.6)
\Line(  57.0,  90.7)(  60.0,  90.7)
\Line(  60.0,  90.0)(  63.0,  90.0)
\Line(  63.0,  89.0)(  66.0,  89.0)
\Line(  66.0,  88.6)(  69.0,  88.6)
\Line(  69.0,  87.8)(  72.0,  87.8)
\Line(  72.0,  86.4)(  75.0,  86.4)
\Line(  75.0,  87.2)(  78.0,  87.2)
\Line(  78.0,  85.6)(  81.0,  85.6)
\Line(  81.0,  85.5)(  84.0,  85.5)
\Line(  84.0,  86.5)(  87.0,  86.5)
\Line(  87.0,  83.6)(  90.0,  83.6)
\Line(  90.0,  84.9)(  93.0,  84.9)
\Line(  93.0,  84.0)(  96.0,  84.0)
\Line(  96.0,  82.1)(  99.0,  82.1)
\Line(  99.0,  81.7)( 102.0,  81.7)
\Line( 102.0,  82.4)( 105.0,  82.4)
\Line( 105.0,  82.1)( 108.0,  82.1)
\Line( 108.0,  81.2)( 111.0,  81.2)
\Line( 111.0,  82.3)( 114.0,  82.3)
\Line( 114.0,  80.4)( 117.0,  80.4)
\Line( 117.0,  81.0)( 120.0,  81.0)
\Line( 120.0,  82.1)( 123.0,  82.1)
\Line( 123.0,  81.7)( 126.0,  81.7)
\Line( 126.0,  80.6)( 129.0,  80.6)
\Line( 129.0,  80.9)( 132.0,  80.9)
\Line( 132.0,  79.8)( 135.0,  79.8)
\Line( 135.0,  80.5)( 138.0,  80.5)
\Line( 138.0,  82.6)( 141.0,  82.6)
\Line( 141.0,  79.9)( 144.0,  79.9)
\Line( 144.0,  79.7)( 147.0,  79.7)
\Line( 147.0,  80.4)( 150.0,  80.4)
\Line( 150.0,  81.7)( 153.0,  81.7)
\Line( 153.0,  80.9)( 156.0,  80.9)
\Line( 156.0,  81.0)( 159.0,  81.0)
\Line( 159.0,  80.8)( 162.0,  80.8)
\Line( 162.0,  79.0)( 165.0,  79.0)
\Line( 165.0,  81.1)( 168.0,  81.1)
\Line( 168.0,  81.4)( 171.0,  81.4)
\Line( 171.0,  80.5)( 174.0,  80.5)
\Line( 174.0,  81.2)( 177.0,  81.2)
\Line( 177.0,  81.4)( 180.0,  81.4)
\Line( 180.0,  82.8)( 183.0,  82.8)
\Line( 183.0,  80.9)( 186.0,  80.9)
\Line( 186.0,  80.4)( 189.0,  80.4)
\Line( 189.0,  81.7)( 192.0,  81.7)
\Line( 192.0,  83.4)( 195.0,  83.4)
\Line( 195.0,  81.7)( 198.0,  81.7)
\Line( 198.0,  84.7)( 201.0,  84.7)
\Line( 201.0,  81.9)( 204.0,  81.9)
\Line( 204.0,  84.1)( 207.0,  84.1)
\Line( 207.0,  82.7)( 210.0,  82.7)
\Line( 210.0,  83.7)( 213.0,  83.7)
\Line( 213.0,  85.5)( 216.0,  85.5)
\Line( 216.0,  84.7)( 219.0,  84.7)
\Line( 219.0,  86.1)( 222.0,  86.1)
\Line( 222.0,  86.5)( 225.0,  86.5)
\Line( 225.0,  87.7)( 228.0,  87.7)
\Line( 228.0,  87.3)( 231.0,  87.3)
\Line( 231.0,  89.2)( 234.0,  89.2)
\Line( 234.0,  89.5)( 237.0,  89.5)
\Line( 237.0,  88.3)( 240.0,  88.3)
\Line( 240.0,  89.7)( 243.0,  89.7)
\Line( 243.0,  91.2)( 246.0,  91.2)
\Line( 246.0,  92.6)( 249.0,  92.6)
\Line( 249.0,  93.4)( 252.0,  93.4)
\Line( 252.0,  94.5)( 255.0,  94.5)
\Line( 255.0,  94.0)( 258.0,  94.0)
\Line( 258.0,  96.6)( 261.0,  96.6)
\Line( 261.0,  98.2)( 264.0,  98.2)
\Line( 264.0,  99.7)( 267.0,  99.7)
\Line( 267.0, 101.0)( 270.0, 101.0)
\Line( 270.0, 104.3)( 273.0, 104.3)
\Line( 273.0, 105.1)( 276.0, 105.1)
\Line( 276.0, 108.6)( 279.0, 108.6)
\Line( 279.0, 111.6)( 282.0, 111.6)
\Line( 282.0, 115.1)( 285.0, 115.1)
\Line( 285.0, 118.7)( 288.0, 118.7)
\Line( 288.0, 123.7)( 291.0, 123.7)
\Line( 291.0, 130.9)( 294.0, 130.9)
\Line( 294.0, 142.5)( 297.0, 142.5)
\Line( 297.0, 215.5)( 300.0, 215.5)
\Line(   3.0, 215.4)(   3.0, 142.7)
\Line(   6.0, 142.7)(   6.0, 130.9)
\Line(   9.0, 130.9)(   9.0, 123.8)
\Line(  12.0, 123.8)(  12.0, 119.0)
\Line(  15.0, 119.0)(  15.0, 114.9)
\Line(  18.0, 114.9)(  18.0, 111.2)
\Line(  21.0, 111.2)(  21.0, 107.4)
\Line(  24.0, 107.4)(  24.0, 106.8)
\Line(  27.0, 106.8)(  27.0, 103.4)
\Line(  30.0, 103.4)(  30.0, 101.2)
\Line(  33.0, 101.2)(  33.0,  99.9)
\Line(  36.0,  99.9)(  36.0,  98.8)
\Line(  39.0,  98.8)(  39.0,  95.8)
\Line(  42.0,  95.8)(  42.0,  95.0)
\Line(  45.0,  95.0)(  45.0,  94.4)
\Line(  48.0,  94.4)(  48.0,  93.5)
\Line(  51.0,  93.5)(  51.0,  92.5)
\Line(  54.0,  92.5)(  54.0,  91.6)
\Line(  57.0,  91.6)(  57.0,  90.7)
\Line(  60.0,  90.7)(  60.0,  90.0)
\Line(  63.0,  90.0)(  63.0,  89.0)
\Line(  66.0,  89.0)(  66.0,  88.6)
\Line(  69.0,  88.6)(  69.0,  87.8)
\Line(  72.0,  87.8)(  72.0,  86.4)
\Line(  75.0,  86.4)(  75.0,  87.2)
\Line(  78.0,  87.2)(  78.0,  85.6)
\Line(  81.0,  85.6)(  81.0,  85.5)
\Line(  84.0,  85.5)(  84.0,  86.5)
\Line(  87.0,  86.5)(  87.0,  83.6)
\Line(  90.0,  83.6)(  90.0,  84.9)
\Line(  93.0,  84.9)(  93.0,  84.0)
\Line(  96.0,  84.0)(  96.0,  82.1)
\Line(  99.0,  82.1)(  99.0,  81.7)
\Line( 102.0,  81.7)( 102.0,  82.4)
\Line( 105.0,  82.4)( 105.0,  82.1)
\Line( 108.0,  82.1)( 108.0,  81.2)
\Line( 111.0,  81.2)( 111.0,  82.3)
\Line( 114.0,  82.3)( 114.0,  80.4)
\Line( 117.0,  80.4)( 117.0,  81.0)
\Line( 120.0,  81.0)( 120.0,  82.1)
\Line( 123.0,  82.1)( 123.0,  81.7)
\Line( 126.0,  81.7)( 126.0,  80.6)
\Line( 129.0,  80.6)( 129.0,  80.9)
\Line( 132.0,  80.9)( 132.0,  79.8)
\Line( 135.0,  79.8)( 135.0,  80.5)
\Line( 138.0,  80.5)( 138.0,  82.6)
\Line( 141.0,  82.6)( 141.0,  79.9)
\Line( 144.0,  79.9)( 144.0,  79.7)
\Line( 147.0,  79.7)( 147.0,  80.4)
\Line( 150.0,  80.4)( 150.0,  81.7)
\Line( 153.0,  81.7)( 153.0,  80.9)
\Line( 156.0,  80.9)( 156.0,  81.0)
\Line( 159.0,  81.0)( 159.0,  80.8)
\Line( 162.0,  80.8)( 162.0,  79.0)
\Line( 165.0,  79.0)( 165.0,  81.1)
\Line( 168.0,  81.1)( 168.0,  81.4)
\Line( 171.0,  81.4)( 171.0,  80.5)
\Line( 174.0,  80.5)( 174.0,  81.2)
\Line( 177.0,  81.2)( 177.0,  81.4)
\Line( 180.0,  81.4)( 180.0,  82.8)
\Line( 183.0,  82.8)( 183.0,  80.9)
\Line( 186.0,  80.9)( 186.0,  80.4)
\Line( 189.0,  80.4)( 189.0,  81.7)
\Line( 192.0,  81.7)( 192.0,  83.4)
\Line( 195.0,  83.4)( 195.0,  81.7)
\Line( 198.0,  81.7)( 198.0,  84.7)
\Line( 201.0,  84.7)( 201.0,  81.9)
\Line( 204.0,  81.9)( 204.0,  84.1)
\Line( 207.0,  84.1)( 207.0,  82.7)
\Line( 210.0,  82.7)( 210.0,  83.7)
\Line( 213.0,  83.7)( 213.0,  85.5)
\Line( 216.0,  85.5)( 216.0,  84.7)
\Line( 219.0,  84.7)( 219.0,  86.1)
\Line( 222.0,  86.1)( 222.0,  86.5)
\Line( 225.0,  86.5)( 225.0,  87.7)
\Line( 228.0,  87.7)( 228.0,  87.3)
\Line( 231.0,  87.3)( 231.0,  89.2)
\Line( 234.0,  89.2)( 234.0,  89.5)
\Line( 237.0,  89.5)( 237.0,  88.3)
\Line( 240.0,  88.3)( 240.0,  89.7)
\Line( 243.0,  89.7)( 243.0,  91.2)
\Line( 246.0,  91.2)( 246.0,  92.6)
\Line( 249.0,  92.6)( 249.0,  93.4)
\Line( 252.0,  93.4)( 252.0,  94.5)
\Line( 255.0,  94.5)( 255.0,  94.0)
\Line( 258.0,  94.0)( 258.0,  96.6)
\Line( 261.0,  96.6)( 261.0,  98.2)
\Line( 264.0,  98.2)( 264.0,  99.7)
\Line( 267.0,  99.7)( 267.0, 101.0)
\Line( 270.0, 101.0)( 270.0, 104.3)
\Line( 273.0, 104.3)( 273.0, 105.1)
\Line( 276.0, 105.1)( 276.0, 108.6)
\Line( 279.0, 108.6)( 279.0, 111.6)
\Line( 282.0, 111.6)( 282.0, 115.1)
\Line( 285.0, 115.1)( 285.0, 118.7)
\Line( 288.0, 118.7)( 288.0, 123.7)
\Line( 291.0, 123.7)( 291.0, 130.9)
\Line( 294.0, 130.9)( 294.0, 142.5)
\Line( 297.0, 142.5)( 297.0, 215.5)
\end{picture}
\end{center}
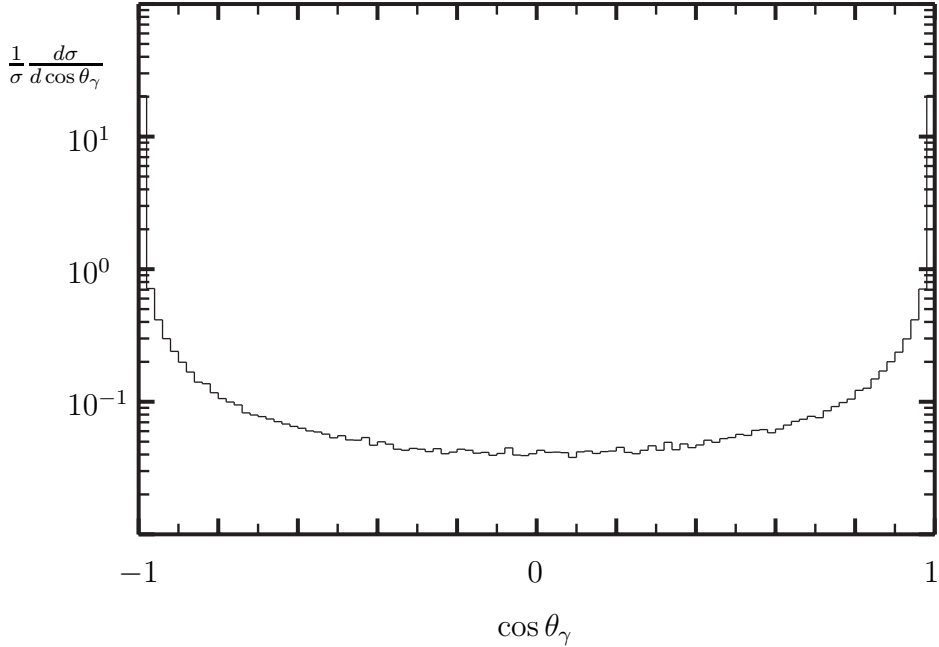
{\em Fig 2: $\cos \theta_\gamma$ distribution (with respect 
to the incoming $e^+$) for the most energetic photon
in the process $e^+ e^- \to \mu^- \bar \nu_\mu u \bar d (\gamma)$.}

\begin{center}
\begin{picture}(400,350)(-50,-50)
\LinAxis(0,0)(300,0)(5,2,5,0,1.5)
\LinAxis(0,300)(300,300)(5,2,-5,0,1.5)
\LogAxis(0,0)(0,300)(6,-5,0,1.5)
\LogAxis(300,0)(300,300)(6,5,0,1.5)
\Text(60 ,-10)[t]{$10$}
\Text(120,-10)[t]{$20$}
\Text(180,-10)[t]{$30$}
\Text(240,-10)[t]{$40$}
\Text(-27, 50)[l]{$10^{-4}$}
\Text(-27,100)[l]{$10^{-3}$}
\Text(-27,150)[l]{$10^{-2}$}
\Text(-27,200)[l]{$10^{-1}$}
\Text(-27,250)[l]{$10^{0}$}
\Text(-50,277)[l]{$\frac{1}{\sigma} \frac{d\sigma}{d E_\gamma}$}
\Text(150,-30)[t]{$E_\gamma$ {\tt [GeV]}}
\Line(   0.0, 255.6)(   3.0, 255.6)
\Line(   3.0, 201.2)(   6.0, 201.2)
\Line(   6.0, 191.1)(   9.0, 191.1)
\Line(   9.0, 184.4)(  12.0, 184.4)
\Line(  12.0, 179.3)(  15.0, 179.3)
\Line(  15.0, 175.1)(  18.0, 175.1)
\Line(  18.0, 171.8)(  21.0, 171.8)
\Line(  21.0, 169.0)(  24.0, 169.0)
\Line(  24.0, 166.6)(  27.0, 166.6)
\Line(  27.0, 163.8)(  30.0, 163.8)
\Line(  30.0, 162.0)(  33.0, 162.0)
\Line(  33.0, 160.1)(  36.0, 160.1)
\Line(  36.0, 158.6)(  39.0, 158.6)
\Line(  39.0, 157.3)(  42.0, 157.3)
\Line(  42.0, 155.0)(  45.0, 155.0)
\Line(  45.0, 154.4)(  48.0, 154.4)
\Line(  48.0, 152.7)(  51.0, 152.7)
\Line(  51.0, 151.2)(  54.0, 151.2)
\Line(  54.0, 149.7)(  57.0, 149.7)
\Line(  57.0, 149.2)(  60.0, 149.2)
\Line(  60.0, 147.9)(  63.0, 147.9)
\Line(  63.0, 147.1)(  66.0, 147.1)
\Line(  66.0, 145.6)(  69.0, 145.6)
\Line(  69.0, 144.4)(  72.0, 144.4)
\Line(  72.0, 144.0)(  75.0, 144.0)
\Line(  75.0, 143.3)(  78.0, 143.3)
\Line(  78.0, 141.9)(  81.0, 141.9)
\Line(  81.0, 141.3)(  84.0, 141.3)
\Line(  84.0, 140.2)(  87.0, 140.2)
\Line(  87.0, 139.0)(  90.0, 139.0)
\Line(  90.0, 138.3)(  93.0, 138.3)
\Line(  93.0, 137.5)(  96.0, 137.5)
\Line(  96.0, 137.2)(  99.0, 137.2)
\Line(  99.0, 136.2)( 102.0, 136.2)
\Line( 102.0, 135.1)( 105.0, 135.1)
\Line( 105.0, 134.2)( 108.0, 134.2)
\Line( 108.0, 134.3)( 111.0, 134.3)
\Line( 111.0, 132.3)( 114.0, 132.3)
\Line( 114.0, 131.9)( 117.0, 131.9)
\Line( 117.0, 130.2)( 120.0, 130.2)
\Line( 120.0, 129.6)( 123.0, 129.6)
\Line( 123.0, 129.7)( 126.0, 129.7)
\Line( 126.0, 129.1)( 129.0, 129.1)
\Line( 129.0, 128.5)( 132.0, 128.5)
\Line( 132.0, 127.2)( 135.0, 127.2)
\Line( 135.0, 126.7)( 138.0, 126.7)
\Line( 138.0, 125.7)( 141.0, 125.7)
\Line( 141.0, 125.2)( 144.0, 125.2)
\Line( 144.0, 123.9)( 147.0, 123.9)
\Line( 147.0, 123.3)( 150.0, 123.3)
\Line( 150.0, 122.0)( 153.0, 122.0)
\Line( 153.0, 121.4)( 156.0, 121.4)
\Line( 156.0, 120.8)( 159.0, 120.8)
\Line( 159.0, 119.5)( 162.0, 119.5)
\Line( 162.0, 118.0)( 165.0, 118.0)
\Line( 165.0, 116.9)( 168.0, 116.9)
\Line( 168.0, 116.3)( 171.0, 116.3)
\Line( 171.0, 115.0)( 174.0, 115.0)
\Line( 174.0, 113.5)( 177.0, 113.5)
\Line( 177.0, 112.2)( 180.0, 112.2)
\Line( 180.0, 110.8)( 183.0, 110.8)
\Line( 183.0, 109.5)( 186.0, 109.5)
\Line( 186.0, 106.8)( 189.0, 106.8)
\Line( 189.0, 106.7)( 192.0, 106.7)
\Line( 192.0, 103.4)( 195.0, 103.4)
\Line( 195.0, 101.7)( 198.0, 101.7)
\Line( 198.0,  99.5)( 201.0,  99.5)
\Line( 201.0,  96.2)( 204.0,  96.2)
\Line( 204.0,  93.1)( 207.0,  93.1)
\Line( 207.0,  90.1)( 210.0,  90.1)
\Line( 210.0,  83.9)( 213.0,  83.9)
\Line( 213.0,  78.9)( 216.0,  78.9)
\Line( 216.0,  74.1)( 219.0,  74.1)
\Line( 219.0,  70.4)( 222.0,  70.4)
\Line( 222.0,  66.9)( 225.0,  66.9)
\Line( 225.0,  62.1)( 228.0,  62.1)
\Line( 228.0,  59.1)( 231.0,  59.1)
\Line( 231.0,  55.6)( 234.0,  55.6)
\Line( 234.0,  52.3)( 237.0,  52.3)
\Line( 237.0,  49.7)( 240.0,  49.7)
\Line( 240.0,  46.4)( 243.0,  46.4)
\Line( 243.0,  44.1)( 246.0,  44.1)
\Line( 246.0,  42.3)( 249.0,  42.3)
\Line( 249.0,  39.5)( 252.0,  39.5)
\Line( 252.0,  38.5)( 255.0,  38.5)
\Line( 255.0,  35.5)( 258.0,  35.5)
\Line( 258.0,  33.5)( 261.0,  33.5)
\Line( 261.0,  31.4)( 264.0,  31.4)
\Line( 264.0,  30.4)( 267.0,  30.4)
\Line( 267.0,  26.4)( 270.0,  26.4)
\Line( 270.0,  25.6)( 273.0,  25.6)
\Line( 273.0,  23.4)( 276.0,  23.4)
\Line( 276.0,  22.3)( 279.0,  22.3)
\Line( 279.0,  20.4)( 282.0,  20.4)
\Line( 282.0,  16.7)( 285.0,  16.7)
\Line( 285.0,  16.2)( 288.0,  16.2)
\Line( 288.0,  15.5)( 291.0,  15.5)
\Line( 291.0,  13.1)( 294.0,  13.1)
\Line( 294.0,  11.6)( 297.0,  11.6)
\Line( 297.0,  11.0)( 300.0,  11.0)
\Line(   3.0, 255.6)(   3.0, 201.2)
\Line(   6.0, 201.2)(   6.0, 191.1)
\Line(   9.0, 191.1)(   9.0, 184.4)
\Line(  12.0, 184.4)(  12.0, 179.3)
\Line(  15.0, 179.3)(  15.0, 175.1)
\Line(  18.0, 175.1)(  18.0, 171.8)
\Line(  21.0, 171.8)(  21.0, 169.0)
\Line(  24.0, 169.0)(  24.0, 166.6)
\Line(  27.0, 166.6)(  27.0, 163.8)
\Line(  30.0, 163.8)(  30.0, 162.0)
\Line(  33.0, 162.0)(  33.0, 160.1)
\Line(  36.0, 160.1)(  36.0, 158.6)
\Line(  39.0, 158.6)(  39.0, 157.3)
\Line(  42.0, 157.3)(  42.0, 155.0)
\Line(  45.0, 155.0)(  45.0, 154.4)
\Line(  48.0, 154.4)(  48.0, 152.7)
\Line(  51.0, 152.7)(  51.0, 151.2)
\Line(  54.0, 151.2)(  54.0, 149.7)
\Line(  57.0, 149.7)(  57.0, 149.2)
\Line(  60.0, 149.2)(  60.0, 147.9)
\Line(  63.0, 147.9)(  63.0, 147.1)
\Line(  66.0, 147.1)(  66.0, 145.6)
\Line(  69.0, 145.6)(  69.0, 144.4)
\Line(  72.0, 144.4)(  72.0, 144.0)
\Line(  75.0, 144.0)(  75.0, 143.3)
\Line(  78.0, 143.3)(  78.0, 141.9)
\Line(  81.0, 141.9)(  81.0, 141.3)
\Line(  84.0, 141.3)(  84.0, 140.2)
\Line(  87.0, 140.2)(  87.0, 139.0)
\Line(  90.0, 139.0)(  90.0, 138.3)
\Line(  93.0, 138.3)(  93.0, 137.5)
\Line(  96.0, 137.5)(  96.0, 137.2)
\Line(  99.0, 137.2)(  99.0, 136.2)
\Line( 102.0, 136.2)( 102.0, 135.1)
\Line( 105.0, 135.1)( 105.0, 134.2)
\Line( 108.0, 134.2)( 108.0, 134.3)
\Line( 111.0, 134.3)( 111.0, 132.3)
\Line( 114.0, 132.3)( 114.0, 131.9)
\Line( 117.0, 131.9)( 117.0, 130.2)
\Line( 120.0, 130.2)( 120.0, 129.6)
\Line( 123.0, 129.6)( 123.0, 129.7)
\Line( 126.0, 129.7)( 126.0, 129.1)
\Line( 129.0, 129.1)( 129.0, 128.5)
\Line( 132.0, 128.5)( 132.0, 127.2)
\Line( 135.0, 127.2)( 135.0, 126.7)
\Line( 138.0, 126.7)( 138.0, 125.7)
\Line( 141.0, 125.7)( 141.0, 125.2)
\Line( 144.0, 125.2)( 144.0, 123.9)
\Line( 147.0, 123.9)( 147.0, 123.3)
\Line( 150.0, 123.3)( 150.0, 122.0)
\Line( 153.0, 122.0)( 153.0, 121.4)
\Line( 156.0, 121.4)( 156.0, 120.8)
\Line( 159.0, 120.8)( 159.0, 119.5)
\Line( 162.0, 119.5)( 162.0, 118.0)
\Line( 165.0, 118.0)( 165.0, 116.9)
\Line( 168.0, 116.9)( 168.0, 116.3)
\Line( 171.0, 116.3)( 171.0, 115.0)
\Line( 174.0, 115.0)( 174.0, 113.5)
\Line( 177.0, 113.5)( 177.0, 112.2)
\Line( 180.0, 112.2)( 180.0, 110.8)
\Line( 183.0, 110.8)( 183.0, 109.5)
\Line( 186.0, 109.5)( 186.0, 106.8)
\Line( 189.0, 106.8)( 189.0, 106.7)
\Line( 192.0, 106.7)( 192.0, 103.4)
\Line( 195.0, 103.4)( 195.0, 101.7)
\Line( 198.0, 101.7)( 198.0,  99.5)
\Line( 201.0,  99.5)( 201.0,  96.2)
\Line( 204.0,  96.2)( 204.0,  93.1)
\Line( 207.0,  93.1)( 207.0,  90.1)
\Line( 210.0,  90.1)( 210.0,  83.9)
\Line( 213.0,  83.9)( 213.0,  78.9)
\Line( 216.0,  78.9)( 216.0,  74.1)
\Line( 219.0,  74.1)( 219.0,  70.4)
\Line( 222.0,  70.4)( 222.0,  66.9)
\Line( 225.0,  66.9)( 225.0,  62.1)
\Line( 228.0,  62.1)( 228.0,  59.1)
\Line( 231.0,  59.1)( 231.0,  55.6)
\Line( 234.0,  55.6)( 234.0,  52.3)
\Line( 237.0,  52.3)( 237.0,  49.7)
\Line( 240.0,  49.7)( 240.0,  46.4)
\Line( 243.0,  46.4)( 243.0,  44.1)
\Line( 246.0,  44.1)( 246.0,  42.3)
\Line( 249.0,  42.3)( 249.0,  39.5)
\Line( 252.0,  39.5)( 252.0,  38.5)
\Line( 255.0,  38.5)( 255.0,  35.5)
\Line( 258.0,  35.5)( 258.0,  33.5)
\Line( 261.0,  33.5)( 261.0,  31.4)
\Line( 264.0,  31.4)( 264.0,  30.4)
\Line( 267.0,  30.4)( 267.0,  26.4)
\Line( 270.0,  26.4)( 270.0,  25.6)
\Line( 273.0,  25.6)( 273.0,  23.4)
\Line( 276.0,  23.4)( 276.0,  22.3)
\Line( 279.0,  22.3)( 279.0,  20.4)
\Line( 282.0,  20.4)( 282.0,  16.7)
\Line( 285.0,  16.7)( 285.0,  16.2)
\Line( 288.0,  16.2)( 288.0,  15.5)
\Line( 291.0,  15.5)( 291.0,  13.1)
\Line( 294.0,  13.1)( 294.0,  11.6)
\Line( 297.0,  11.6)( 297.0,  11.0)
\end{picture}
\end{center}

\vspace{-0.5cm}
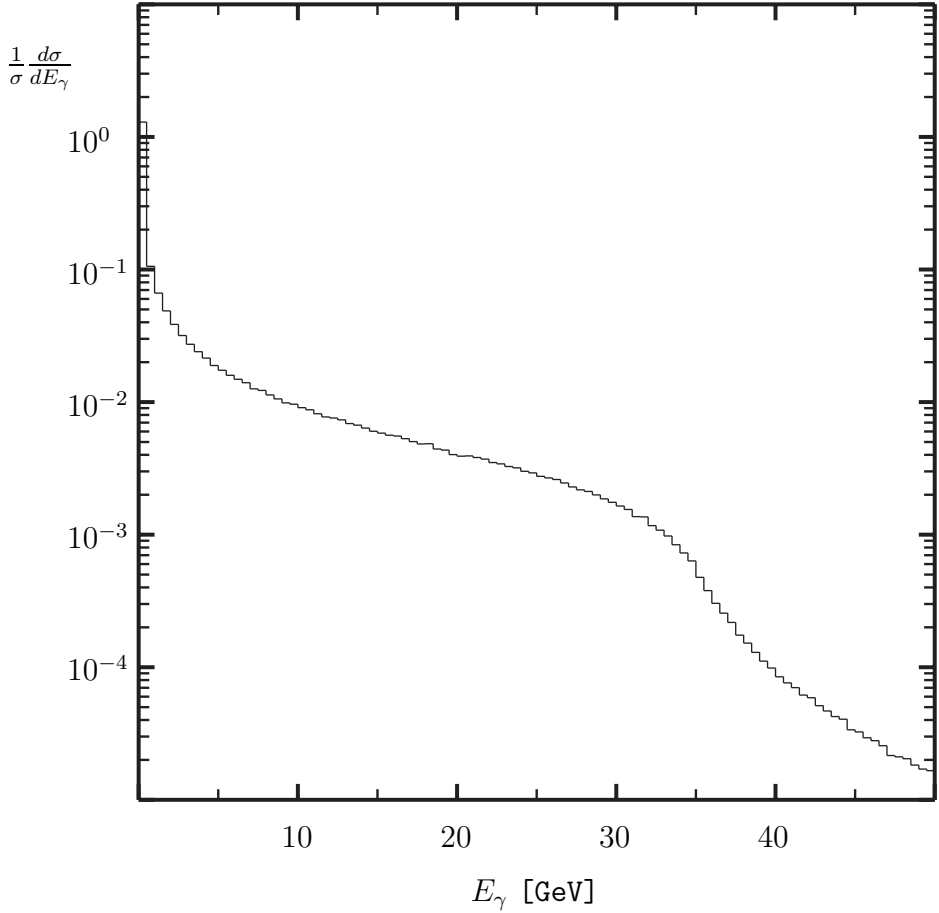
\noindent {\em Fig 3: $E_\gamma$ distribution for the most energetic photon
in the process\\ $e^+ e^- \to \mu^- \bar \nu_\mu u \bar d (\gamma)$.}

\vspace{0.6cm}

Finally, in tables 8 and 9, we show comparisons with the Higgs cross sections
published in ref. \cite{lep2h}, by choosing {\tt WPHACT} \cite{wphact} 
as a benchmark program. 
We devoted special care to implement exactly the same input parameters
of ref. \cite{lep2h}.
For completeness we list them here:
\begin{itemize}
\item Standard LEP2 input parameter set (see ref. \cite{lep2}). 
\item Massless fermions everywhere, except in the Higgs coupling to the $b$.
\item Running widths in the bosonic propagators.
\item $\Gamma_H= \frac{\alpha\,m_H}{8\,M_W^2\,\sin^2_{\theta_W}}(
 m^2_\tau+3 m_b^2 +3 m_c^2)$. 
\item $m_\tau= 1.777$ GeV, $m_b= 2.9$ GeV and $m_c= 0.75$ GeV.
\item No ISR, no QCD corrections but all background diagrams included.
\item $M_Z$ - 25 GeV $\le m_{\ell \bar \ell} \le M_Z$ + 25 GeV and
      $ m_{b \bar b} \ge $ 50 GeV.
\end{itemize}
It is worth mentioning explicitly that {\tt NEXTCALIBUR} can 
consistently include fermion masses everywhere, and that 
they have been neglected in the presented numbers
just for the sake of comparison.

\section{Conclusions}
We introduced {\tt NEXTCALIBUR}, a new Monte Carlo program 
to study four-fermion processes in $e^+e^-$ collisions. 
We outlined our strategy for including Higgs, fermion masses and
leading higher order effects, without loosing efficiency
in the event generation.
The program is meant to upgrade the performances of
an already existing code \cite{exca}.

\begin{figure}
\begin{center}
\begin{tabular}{|l||c|c|c|} \hline 
$m_H$ (GeV)       & 65 & 90 & 115      \\ \hline \hline
 Final state & \multicolumn{3}{c|}{$\mu^+ \mu^- b \bar b$}  \\ \hline
{\tt   WPHACT}    &32.7141(68)&1.59946(64)&1.05953(56) \\ \hline 
{\tt NEXTCALIBUR} &32.691(19) &1.5999(18) &1.0588(11)  \\ \hline 
 Final state & \multicolumn{3}{c|}{$\nu_\mu \bar \nu_\mu b \bar b$} 
                                  \\ \hline \hline
{\tt   WPHACT}    &64.238(14)&2.3661(10)&1.29237(82)\\ \hline 
{\tt NEXTCALIBUR} &64.256(35)&2.3651(27)&1.2910(14) \\ \hline 
 Final state & \multicolumn{3}{c|}{$\nu_e \bar \nu_e b \bar b$}  
                                  \\ \hline \hline
{\tt   WPHACT}    &71.694(27)&5.0996(23)&1.08027(89)\\ \hline 
{\tt NEXTCALIBUR} &71.778(84)&5.086(12) &1.0776(13) \\ \hline 
\end{tabular}                 
\end{center}
{\em Table 8: Cross sections, in fb, for Higgs production
at $\sqrt{s}$= 175 GeV.

\noindent See text for the input parameters.}
\end{figure}
\begin{figure}
\begin{center}
\begin{tabular}{|l||c|c|c|} \hline 
$m_H$ (GeV)       & 65 & 90 & 115      \\ \hline \hline
 Final state & \multicolumn{3}{c|}{$\mu^+ \mu^- b \bar b$}  \\ \hline
{\tt   WPHACT}    &37.3990(64)&24.4727(40)&10.7027(24)\\ \hline 
{\tt NEXTCALIBUR} &37.394(21) &24.471(14) &10.7006(77)\\ \hline 
 Final state & \multicolumn{3}{c|}{$\nu_\mu \bar \nu_\mu b \bar b$} 
                                  \\ \hline \hline
{\tt   WPHACT}    &72.927(16)&47.222(12)&19.841(11)\\ \hline 
{\tt NEXTCALIBUR} &72.929(46)&47.231(33)&19.842(19)\\ \hline 
 Final state & \multicolumn{3}{c|}{$\nu_e \bar \nu_e b \bar b$}  
                                  \\ \hline \hline
{\tt   WPHACT}    &80.611(34)&53.335(19)&20.893(12)\\ \hline 
{\tt NEXTCALIBUR} &80.507(96)&53.280(67)&20.897(24)\\ \hline 
\end{tabular}                 
\end{center}
{\em Table 9: Cross sections, in fb, for Higgs production 
at $\sqrt{s}$= 192 GeV.

\noindent See text for the input parameters.}
\end{figure}

We concentrated mainly on QED and scale-dependent corrections,
without making any attempt to include genuine weak contributions.
While the latter are certainly relevant for LEP2 precision measurements,
such as $\sigma_{WW}$ and $M_W$, they do not seem to be necessary for all 
the other observables. In that respect {\tt NEXTCALIBUR} represents
a solid tool for the final analysis of the LEP2 four-fermion 
data and for studying $e^+ e^-$ Physics in general.

 In the near future, two big improvements of the program are foreseen.
First of all the inclusion in the matrix element of anomalous couplings.
 Secondly, the implementation of the formalism of ref.~\cite{bbc} 
to incorporate running couplings and finite boson widths effects
without breaking $SU(2) \times U(1)$ gauge invariance.

 Notice that, given our computational strategy, the only needed modification
is the insertion of additional Feynamn rules in the matrix element, 
all the rest remaining the same. Therefore, we do not expect 
difficulties of principle in the actual implementation.

\section*{Acknowledgements}
Fruitful discussions with all the participants in the CERN LEP2 Monte
Carlo Workshop \cite{wshop} are acknowledged, 
in particular with Giampiero Passarino and Alessandro Ballestrero.

\end{document}